# ARE THE LUMINOSITIES OF RR LYRAE STARS AFFECTED BY SECOND PARAMETER EFFECTS?

SIDNEY VAN DEN BERGH

Dominion Astrophysical Observatory

5071 West Saanich Road

Victoria, British Columbia

V8X 4M6, Canada

## ABSTRACT

There is a serious discrepancy between the distance to the LMC derived from the Cepheid Period-Luminosity relation and that obtained by using the Galactic calibration for the luminosity of RR Lyrae stars. It is suggested that this problem might be due to the fact that second parameter effects make it inappropriate to apply Galactic calibrations to RR Lyrae variables in the Magellanic Clouds, i.e. $M_V$ (RR) could depend on both [Fe/H] and on one or more second parameters.

## 1.  INTRODUCTION

It has been known for many years (van den Bergh 1965, 1967, Sandage & Wildey 1967) that the population gradients along the horizontal branches of globular clusters depend on metallicity [Fe/H] and on at least one other parameter. This raises the question whether the luminosity of the horizontal branch, and hence $M_V$ (RR), might also be affected by a parameter other than [Fe/H]. It is the purpose of this letter to explore whether the discrepancy between the Cepheid and RR Lyrae distance scales to the Large Magellanic Cloud (Walker 1992), and to some other Local Group galaxies (van den Bergh 1995), might be due to such a second parameter effect. Models of horizontal branch stars suggest that second parameter effects might be due to variations in age, helium abundance Y, [CNO/Fe] or core rotation rate (Lee 1993).

## 2.  THE DISTANCE TO THE LMC

Distance determinations to Local Group galaxies have recently been reviewed by van den Bergh (1995). Cepheids and RR Lyrae stars are found to provide the highest weight methods for determining the distance to the Large Magellanic Cloud. From a comparison of JHK observations of Cepheids in the LMC with similar observations of Galactic calibrators in open clusters and associations, Laney & Stobie (1994) obtain a true distance modulus $(m-M)_o = 18.53$, with an internal standard error of 0.04 mag.

Both the Baade-Wesselink method (Carney, Storm & Jones 1992) and statistical parallaxes of Galactic halo RR Lyrae stars (Layden, Hanson & Hawley 1994, Layden, Hawley & Hanson 1995) are consistent with

$$< M_V (RR) > \ = \ 0.15 \ [Fe/H] \ + \ 1.0, \qquad (1)$$

in which the zero-point uncertainty $\simeq$ 0.08 mag. A much steeper dependence of $M_V (RR)$ on [Fe/H], that has been proposed by Sandage & Cacciari (1990) and by Sandage (1993), is <u>not</u> supported by the observed magnitude difference between the metal-rich cluster K 58 ([Fe/H] = -0.57) and that of the metal-poor cluster K 219 ([Fe/H] = -1.83) in M31 which has been observed by Ajhar et al. (1994). The M31 cluster observations are, however, consistent with Eqn. (1). From application of Eqn. (1) to the 182 RR Lyrae stars that Walker (1992) observed in seven globular clusters having -2.3 $\leq$ [Fe/H] $\leq$ -1.7 that are associated with the LMC one obtains $(m-M)_o = 18.24 \pm 0.09$, in which the quoted error includes the estimated uncertainty in the zero-point of Eqn. (1).

Sandage (1993) finds that

$$< M_V (RR) > = 0.30 [Fe/H] + 0.94. \quad (2)$$

Adoption of this relation would slightly increase the LMC distance modulus derived from Walker's (1992) observations of cluster RR Lyrae stars. From Eqn. (2) one obtains an LMC distance modulus $(m-M)_o = 19.31 \pm 0.05$ (internal error), which would marginally decrease the discrepancy between the Cepheid and RR Lyrae distances to the Large Cloud.

It is noted in passing that the horizontal branch models by Dorman (1992) support the small value of the slope in Eqn. (1), rather than the larger one in Eqn. (2).

Clearly the Cepheid distance modulus $(m-M)_o = 18.53 \pm 0.04$ and the RR Lyrae modulus $(m-M)_o = 18.24 \pm 0.09$ are discrepant. Observations of SN 1987A are not much help in choosing between the two values. From observations of the fluorescence of the ring surrounding this object, Gould (1994a) obtained a distance modulus $(m-M)_o = 18.63 \pm 0.10$, which would be consistent with the Cepheid distance modulus, but not with the modulus derived from RR Lyrae stars.

However, a more recent study by Gould (1995) suggests that $(m-M)_o = 18.37 \pm 0.04$ for a circular ring. The latter value is intermediate between those derived from Cepheids and RR Lyrae stars.

The question whether the difference between the Cepheid and RR Lyrae distances to the LMC could be due to abundance effects on the Cepheid Period-Luminosity relation has been discussed by Gould (1994b) and by van den Bergh (1995). Assuming [Fe/H] = -0.3 (Z = 0.009) for young stars in the Large Cloud (Spite & Spite 1991), in conjunction with $\Delta Y/\Delta Z = 3$, and Y (primordial) = 0.228 (Pagel 1992), the classical evolutional models of Chiosi, Wood & Capitanio (1993) yield a Period-Luminosity relation for the LMC that is only 0.01 mag brighter for 10-day Cepheids than it is for similar objects in the Galaxy. It is therefore tentatively concluded that it is legitimate to apply the Galactic zero-point calibration to classical Cepheids in the Large Cloud. If this conclusion is correct, then the discrepancy between the Cepheid and RR Lyrae distance scales to the Large Cloud might be due to the fact that the Galactic calibration of $M_V$ (RR) is not applicable to the Large Cloud.

## 3. DISCUSSION

In a recent review van den Bergh (1995) found that discrepancies between Cepheid and RR Lyrae distance scales, similar to that observed in the LMC, also occur in the SMC and in IC 1613, but not in M 31. This suggests a possible correlation between second parameter effects on $M_V$ (RR) and parent galaxy luminosity.

Horizontal branch models by Sweigart (1987), Lee (1990) and Dorman (1993) show that the luminosities of RR Lyrae stars increase with rising helium abundance. The effects of other changes, such as variations in [$\alpha$/Fe], e.g. Yi, Lee & Demarque (1993) have not yet been studied in detail.

Adoption of the RR Lyrae distance scale to the LMC would decrease the distance to the Large Cloud, resulting in a 14% <u>increase</u> in the value of the Hubble parameter $H_o$ that is derived from Cepheids in the Virgo cluster (Pierce et al. 1994, Freedman et al. 1994). This in its turn would increase the discrepancy between the Cepheid distance scale and that derived from supernovae of Type Ia (Branch & Khokholov 1995). Furthermore, a short distance scale to the LMC would decrease the ages of the oldest globular clusters from ~ 18 Gyr to only ~ 14 Gyr (Chaboyer 1995).

I am indebted to Andy Layden for useful discussions about the luminosities of RR Lyrae stars, to Ben Dorman for helpful information on models for horizontal branch stars, and to an anonymous referee for numerous helpful suggestions.

- 9 -